# A hackable, multi-functional, and modular extrusion 3D printer for soft materials

Iek Man Lei[1,2], Yaqi Sheng[1,2], Chon Lok Lei[3,4], Cillian Leow[1] and Yan Yan Shery Huang[1,2]*

*1 Department of Engineering, University of Cambridge, Cambridge, United Kingdom*
*2 The Nanoscience Centre, University of Cambridge, Cambridge, United Kingdom*
*3 Institute of Translational Medicine, Faculty of Health Sciences, University of Macau, Macau*
*4 Department of Computer Science, University of Oxford, Oxford, United Kingdom*

*\* Corresponding: yysh2@cam.ac.uk*

### Running title:

To democratize 3D printing for soft materials, the authors constructed Printer.HM, which can perform versatile operations, including liquid dispensing, non-planar printing, and pick-and-place. Printer.HM was used to create pH-responsive soft actuators, and shaping plant-based hydrogels; thus, laying down future open innovations in soft, biological, and sustainable material architectures.

## Acknowledgement

This work was supported by the European Research Council (ERC-StG, 758865). I.M.L. acknowledges the financial support from the W.D. Armstrong Trust and the Macao Postgraduate Scholarship Fund. C.L.L. acknowledges the support from the University of Macau via a UM Macao Fellowship and the Clarendon Scholarship Fund. Y.S. acknowledges the financial support from the Chinese Scholarship Council. The authors thank Dr Yang Cao for her assistance with the cell dispensing experiment, Abby Thompson for her help with the 3D scanner, Ian Ganney for his help with the machining work, and Dr Clement Chun Lam Chan and Prof Silvia Vignolini (Department of Chemistry, University of Cambridge) for providing the methacrylate hydroxypropyl cellulose solution.

## Conflict of Interest

The authors declare no competing interests.

**Abstract**

Three-dimensional (3D) printing has emerged as a powerful tool for material, food, and life science research and development, where the technology's democratization necessitates the advancement of open-source platforms. Herein, we developed a hackable, multi-functional, and modular extrusion 3D printer for soft materials, nicknamed Printer.HM. Multi-printhead modules are established based on a robotic arm for heterogeneous construct creation, where ink printability can be tuned by accessories such as heating and UV modules. Software associated with Printer.HM were designed to accept geometry inputs including computer-aided design models, coordinates, equations, and pictures, to create prints of distinct characteristics. Printer.HM could further perform versatile operations, such as liquid dispensing, non-planar printing, and pick-and-place of meso-objects. By 'mix-and-match' software and hardware settings, Printer.HM demonstrated printing of pH-responsive soft actuators, plant-based functional hydrogels, and organ macro-anatomical models. Integrating affordability and open design, we envisage the Printer.HM concept to widen open innovations for soft, biological, and sustainable material architectures.

# 1. Introduction

The advent of 3D printing offers potential freedoms to rapidly create arbitrarily architected matter from a wide variety and combination of soft and functional materials, revolutionising diverse research fields, from food to tissue engineering, and soft electronics and robotics[1–6]. Among different 3D printing modalities for soft materials, extrusion-based printing is arguably the most widely used modality owing to its broad material compatibility, low material usage, low wastes, and its ability to spatially control the construct properties[7–11], such as composition[12], photonic properties[13], orientation of encapsulated fibres[14,15] and ferromagnetic properties[16]. However, ongoing innovations in extrusion 3D printing need to overcome the cost barrier and the limited adaptability associated existing commercial systems. Although several open-source, custom-made extrusion 3D printers have been reported[17–21], the printable materials and architectures were restricted by an incomplete set of auxiliary tools and print path options (as summarized in Supplementary Table I and Supplementary Figure 1). To democratize multi-functional extrusion printing, we report the development of a multi-printhead, and highly customisable extrusion-based 3D printer for soft materials. We name the printer 'Printer.HM' herein, where HM stands for Hackable and Multi-functional. Printer.HM can readily accept different geometry inputs, such as coordinates, equations, and pictures, in addition to the conventional computer-aided design (CAD)-G-code input. The total costs of establishing a version of Printer.HM are between £900 and £1900 (taking between 2 to 4 hrs of installation time), depending on the number of utilities equipped. Printer.HM offers excellent print compatibility with a wide variety of liquidous and soft materials (from mPa.s to kPa.s); and myriad operations can be performed, including liquid dispensing, multi-material printing, printing with variable speed, non-planar printing and pick-and-place application. By virtue of the modular design of 'Printer.HM' and the use of a robotic arm as the motion control, users can easily assemble and reconfigure the setup, and expand its functionalities, based on individual's experimental needs. Overall, we foresee the hackable, expandable, and affordable nature of Printer.HM can promote the widespread adaptability of extrusion 3D printing technology, facilitating open innovations in the research communities which utilise soft, biological, and sustainable materials.

## 2. Results and discussion

**A modular extrusion printer based on a robotic arm**

Figure 1 highlights the key features of 'Printer.HM', an extrusion-based 3D printer as an affordable and hackable alternative to commercial bioprinters. The system is built on a hackable robotic arm (see Supplementary Figure 2, for a real-life setup), which does not rely on repurposing an existing fused deposition modelling printer, as their proprietary firmware may still limit its customisability and the number of utilities (e.g. printheads and UV module) that can be fitted into the system[20,22]. Housed in an enclosure and built on an aluminium breadboard for ease of reconfiguring different modules, the core part of the printer consists of dispensing modules (i.e. four custom-built piston-driven printheads); and a stage, of which motion control is afforded by the robotic arm. The use of a moveable built plate that is controlled by a robotic arm here, instead of a set of 3-axis Gantry linear stages, offers advantages of compactness and ease of assembly.

As the printing competency of soft materials crucially depends on the rheology and crosslinking of the inks, 'Printer.HM' is equipped with add-on auxiliary tools, including a syringe heater a stage heater, and a UV module for assisting printing of hydrogels (Supplementary Figure 2). The heaters are capable of controlling the temperatures of the stage and the syringe from room temperatures to ~ 60 $^{o}$C, which is sufficient for most types of hydrogels and elastomeric materials. Various stages were custom-designed to fit different sizes of receiving substrates or reservoirs, including standard glass slides, petri dishes (90, 55 and 35 mm) and rectangular containers (40 and 30 mm) (Supplementary Figure 2b). The printheads of 'Printer.HM' were built from simple mechanical components, such as lead screw, micro-stepper motors with dispensing resolution of 0.8 μm per step (see Supplementary Note IV) and linear rails for increased stability and compactness. The syringe holders of the printheads were 3D printed, enabling customisation for fitting different sizes of dispensing utilities specific to the users' experiments. As proof-of-concept, we customised the printheads to accommodate 3 ml or 1 ml syringes that are compatible with most laboratory-based applications.

Being able to freely tailor the print path is important as it directly controls the properties of the printed constructs, such as their mechanical properties[23], stimuli-responsive morphing behaviour[24,25] and the cell orientation in the constructs[26]. Commercial and existing custom-made extrusion-based printers generally employ CAD models/G-code as the only option to describe printing designs[10,27,28]. The lack of geometry input options offered in these systems could restrict the design freedom and the customisability of the print path especially for actuator structures[29]. Thus, printer.HM was designed to accept four different geometry inputs for creating prints of distinct characteristics. They are coordinates, equations, G-codes and pictures.

Overall, the modular design of 'Printer.HM' allows users to reconfigure the setup based on their experimental requirements and resource limitation, as well as encouraging the research community to expand the functionalities of the system via designing new modules. As a proof-of-concept, four printheads were built here. The total cost of this fully equipped and four-printhead system is around ~ £1900, while a single printhead system costs ~ £900. This offers significant cost saving compared to commercial bioprinters[10]. The installation time of 'Printer.HM' is around 2 to 4 hrs, excluding the time required for 3D printing parts. A step-by-step instruction of the printer assembly is provided in Supplementary Note III, to promote the reproducibility of the system.

**Printing competency**

Compared to the existing open-source printers, the broader set of auxiliary tools associated with 'Printer.HM' greatly enhances its ability to construct different materials and geometry combinations (see Supplementary Figure 2 and Supplementary Table 1). To exemplify, Printer.HM was used to perform extrusion printing of soft materials that require different gelation mechanisms, including thermally-induced gelation and photo-induced crosslinking, as shown in Figure 2a. Thermal-responsive hydrogels self-assemble and undergo phase transitions at their critical temperatures that are defined by lower critical solution temperature (LCST) or upper critical solution temperature (UCST)[30,31]. UCST hydrogels, including gelatin and agarose, undergo gel formation upon cooling at a temperature below their UCST[32,33]. On the contrary, gelling of LCST (e.g. Pluronic F127) hydrogels occurs when increasing the temperature above their LCST[33]. With Printer.HM, the syringe heater assists the printing of UCST hydrogels (i.e. gelatin) via heating the inks during extrusion, while the stage heater helps preserving the printed shape of LCST hydrogels via enhancing its rheology at elevated temperature at the built plate. The UV module allows *in-situ* crosslinking of photo-polymerizable hydrogels (i.e. methacrylate hydroxypropyl cellulose[34]) during printing. In addition, as shown by Figure 2b, the printer is capable of printing a wide variety of biomaterials, from Poly(ethylene glycol) diacrylate (PEGDA, a low viscosity ink with viscosity 20 mPa.s[35]), collagen, silicone elastomer to a highly viscous solution of sodium carboxyl methyl cellulose (1500 Pa.s).

We further demonstrate the print resolution of 'Printer.HM' by printing a line pattern with Pluronic F127. The test was performed with Pluronic F127 here as it is commonly used in literature. Using a non-optimised setting of the operation parameters, the median Pluronic F127 feature achieved with 'Printer.HM' was around ~150 μm (Figure 2c), which is comparable with the resolution typically attained in extrusion-based bioprinting[27,36]. However, it should be noted that the resolution of the printed features is predominantly determined by the nozzle size and the ink properties.

**Printing with versatile geometry inputs**

Figure 3 illustrates the wide variety of constructs fabricated using different geometry inputs, each with their own strengths depending on the architecture requirement. Being the most ordinary form of the geometry inputs, coordinates is particularly useful for creating simple linear or regular patterns, such as one-dimensional channels (Figure 3a, Supplementary Video 1). Meanwhile, equation input enables the creation of seamless one-stroke curvy patterns, however it is not suitable for complex patterns that are not describable by equations (Figure 3b, Supplementary Video 2). Figure 3b demonstrates that simple tubular constructs can be readily produced via an equation of circle, without the need to prepare CAD files. On the other hand, three-dimensional intricate objects can be well-described by 3D CAD models, the standard geometry format used in 3D printing (Figure 3c, Supplementary Video 3). Lastly, picture input enables the creation of customised motifs via photos of hand-drawn patterns or pictures created by any drawing software. By leveraging the picture input option, user-designed patterns, for example circuit- and vascular-like patterns, can be readily fabricated (Figure 3d.i, 3d.ii).

To further illustrate the benefits of having print path customisability, in particular for soft robotics applications, we demonstrated the creation of a soft morphing system made of a pH-responsive hydrogel by leveraging an anisotropic print path. As showed in Figure 3d.iii and Supplementary Video 4, the 2D construct created with a heterogeneous print path exhibited an anisotropic swelling response and morphed into a flower shape. It is worthwhile mentioning that the operation flexibility does not limit to the four geometry inputs provided here. As the control programme is entirely hackable, users can freely amend the programme for unprecedented designs.

**Multi-functionalities in one platform**

By virtue of the customisable control programme, operations with 'Printer.HM' are user-amendable and multi-functional. We demonstrate that operations, such as automated dispensing, printing with variable speed and non-planar printing, can be carried out with 'Printer.HM'. To exemplify, liquid handling always plays an indispensable role in life science experiments. Thus, we transformed 'Printer.HM' into a dispenser by modifying the control programme. Figure 4a and Supplementary Video 5 show that droplets of cell suspension were automatically dispensed on a petri dish. The dispensed volume of the droplets is controllable by the extrusion flowrate and the dispensing duration. By simply setting different duration of the dispensing time, droplets with various sizes were obtained. This capability might be useful for automating the hanging drop method to produce cell spheroids and dispensing active ingredients within a printed object.

Features with continuously narrowing width can be easily generated with variable speed of the stage, as demonstrated in Figure 4b, which can assist in creating hierarchical vascular network. Further, we demonstrate the capability of performing non-planar printing using the platform. As opposed to conventional commercial 3D printers that rely on plane-by-plane slicing, non-planar printing requires the ink to be printed on a freeform surface by moving the motion part of the printer in all 3 axes at the same time. Figure 4c and Supplementary Video 6 show a line pattern was printed on a 3D target nose model. The 3D surface of the target object was assessed using a 3D scanner and the line pattern was projected according to the non-planar geometry using a custom-written code (further described in Materials and Methods). With the ability to deposit inks directly onto variable object surfaces, novel applications of 3D extrusion printing technology could become possible, such as depositing freeform circuits[37] and functional materials. Lastly, we demonstrate nozzle-based 'pick-and-place' of meso-objects in Figure 5 and Supplementary Video 7. Such an operation enables applications for locating cell spheroids between different environments.

The multiple printheads equipped in 'Printer.HM' facilitate the fabrication of multi-material constructs. As a demonstration, Figure 6a shows a four-layer construct composed of Pluronic F127 inks coloured with different dyes printed in air, and a model of the respiratory system with lungs and trachea made of alginate inks printed inside a support bath (Figure 6b). This capability opens up future potential in generating sophisticated tissue anatomy, which are usually multi-component and spatially heterogeneous.

## 3. Conclusion

Extrusion 3D printing is a promising approach for fabricating soft tissue constructs and biomimetic soft actuators[10,38]. However, commercial printers could typically be cost-prohibitive, and do not allow ample customisation. These limitations greatly hinder the continuous innovation of the technology and its widespread adoption particularly in resource-limited community[10]. To address these limitations, here, we present an affordable and highly customisable open-source extrusion 3D printer, Printer.HM, that is equipped with multiple printheads, heaters and UV module for soft material printing. The printer was built from simple mechanical components and 3D printed parts that can be readily-sourced and fabricated. A robotic arm was employed as the motion control as it offers the advantages of compactness and ease-of-assembly. Printer.HM offers affordability (£1900 for a four-printhead system), and compatibility with smaller size of syringes that is desirable in small-scale biological applications. Remarkably, the unconventional geometry input options offered in 'Printer.HM' enable the fabrication of prints with distinct design. Using the picture geometry input, users without CAD experience can facilely customise the print path, which is particularly beneficial for controlling the morphing behaviours of stimuli-responsive hydrogels. Despite the low-cost and the custom-made nature of the system, 'Printer.HM' offers good printing competency with a wide variety of soft materials, from hydrogels to silicone elastomers over a wide range of viscosity (20 mPa.s – 1.5 kPa.s). In addition, the system is capable of performing a range of unconventional tasks, such as printing with variable speed and non-planar printing.

Notwithstanding, several limitations are noted in 'Printer.HM'. First, the design of a moving stage system used in 'Printer.HM' might potentially compromise the fidelity of low viscosity objects that are printed in-air. To reduce the potential impact, a very slow speed of the stage can be used when printing delicate structures of low viscosity materials. Second, 'Printer.HM' does not encompass cooling systems that assist printing of protein-based inks and an ink retraction mechanism that prevent inks from unintentional oozing. Nevertheless, the modular design of the system enables easy reconfiguration and the expandability of the system. Users can incorporate new functionalities, such as microfluidic printheads, coolers etc., to 'Printer.HM' in future development. In summary, our work established an affordable 3D extrusion printer with improved customisability and functionalities, benefitting the do-it-yourself research community and potentially facilitating the development of open and innovative fabrication strategies in diverse fields, such as tissue engineering, soft robotics, food, and eco-friendly material processing.

# 4. Materials and methods

**Mechanical design**

'Printer.HM' is an open-source extrusion 3D printer that consists of a commercially available open-source robotic arm (uArm Swift Pro Desktop Robotic Arm) and a dispensing module as the core part, and heating systems, a UV module and an inspection camera as optional utilities. The robotic arm controlled the x, y and z axis motion of the 3D printed stage. Various stages were custom-designed to fit different sizes of receiving substrates or reservoirs, including standard glass slides, petri dishes (90, 55 and 35 mm) and rectangular containers (40 and 30 mm) (Supplementary Figure 2b). The dispensing module is composed of do-it-yourself (DIY) piston-driven printheads that were built from simple mechanical components (i.e. stepper motor, linear rail and ball bearing) and custom-designed 3D printed parts. All CAD files of the 3D printed parts of 'Printer.HM' are accessible and will be made available on Github, thus users can freely amend the parts to better tailor to their applications if needed. The 3D printed parts were printed with polylactic acid (PLA) or acrylonitrile butadiene styrene (ABS) using an Ultimaker S3 3D printer. As a proof-of-concept, four printheads were built here and they were designed to accommodate 1 ml or 3 ml syringes, but users can adjust the number of printheads or amend the design of the syringe holder to fit other sizes of dispensing tools in accordance with their experimental need.

The stage and syringe heating systems in 'Printer.HM' are composed of a custom-made aluminium holder that was wrapped with nichrome wires (UMNICWIRE2, Ultimachine) as the heating element and a K-type thermocouple (Z2-K-1M, Labfacility) as the temperature sensor. A UV LED light source (5 W, 365 nm, NSUV365, Nightsearcher) was employed here and was mounted onto the aluminium breadboard of 'Printer.HM'. Meanwhile, users can select different light sources based on the choice of the photo-initiators. An inspection camera unit was mounted onto the aluminium breadboard for *in-situ* monitoring and recording the printing process. The dispensing module and the heating systems were connected to Arduino boards, while the robotic arm has a built-in Arduino for controlling. Assembly instruction of the printer and the electrical circuit of 'Printer.HM' is described in Supplementary Note III.

**Programme description**

The printing operation was implemented by a custom-written Python programme that synchronously communicates with the Arduino boards of the robotic arm and the dispensing module, whereas the heating modules were independently controlled by graphical user interfaces (GUI) that communicate with the Arduino boards of the heaters which users can freely customise the programme for their needs. All the operation programme used in this study will be made available on Github.

**Printing operation**

Prior to printing, the ink was centrifuged at 1000 g for 3 mins to remove bubbles. The ink was drawn into a 1 ml or 3 ml syringe, and the syringe was loaded to the syringe holder of the setup. A collecting reservoir, such as petri dish or glass slides, was loaded to the 3D printed custom-made stage. Four Python control programmes were written for importing different types of geometry inputs – coordinates, equation, CAD model and picture inputs. Printing parameters, such as printing speed, offset position, extrusion flowrate and initial z-position, are user-adjustable and can be defined in the control programme. By default, the constructs were printed at the centre of the collecting reservoir, unless an offset position was defined.

*Printing with coordinate input*

A list of coordinates ($x = [x_1, x_2, \ldots, x_n]$, $y = [y_1, y_2, \ldots, y_n]$) was directly loaded to the programme, where $x_n$ and $y_n$ denote the $x$ and $y$ coordinates of the $n^{th}$ point (see Supplementary Figure 8).

*Printing with equation input*

A set of cartesian or parametric equations together with the defined range of the independent variable was inputted in the control programme (see Supplementary Figure 9). The curve was discretised by at least 100 evenly spaced points, depending on the length of the curve. The constructs shown in Figure 3b were fabricated using equations of sine wave, butterfly curve and circle. 3D features were produced by printing stacked layers of the 2D curve according to the defined object and layer heights.

*Printing with CAD model input*

3D CAD models were either designed using Autodesk Inventor or downloaded from GradCAD (https://grabcad.com/library/software/stl) or Thingiverse (www.thingiverse.com). Prior to the printing process, the CAD model was converted to a G-code file using Slic3r (https://slic3r.org/) with the user-defined slicing parameters (i.e. fill pattern, fill density, extrusion width and layer height). The G-code file was then imported to the Python control programme.

*Printing with picture input*

Pictures of the printing design or photos of the hand-drawn sketches were imported to Inkscape. They were converted to G-code using the 'Gcodetools' extension on Inkscape (https://inkscape.org/), which was an extension designed for CNC machines. Step-by-step procedure of the conversion can be found in Supplementary Note VI. The generated G-code was then imported to the control programme for picture input, which was written to accept the G-code generated by this extension.

**Heating operation**

Syringe heating and stage heating were applied when required. They were controlled by a custom-written graphical user interface (GUI), where users can directly specify the desired set-point temperature. The acceptable deviation from the desired set-point temperature was defaulted to ± 0.5 $^{o}$C here. The control programme for the heating operation will be made available on Github.

**Non-planar printing**

A 2D line pattern for printing was designed on Inkscape and was converted to a G-code file. The 3D shape of the target object (a nose model made of Ecoflex) was captured using a 3D scanner (EinScan H, SHINING 3D®) and was saved as a STL file. To analyse the surface of the target object, the STL file of the nose model was converted into a G-code file using Slic3R with the following slicing settings (fill pattern = 'Hilbert curve', extrusion width = 0.2 mm, fill density = 100% and layer height = 0.2 mm). A dense infill setting and a Hilbert curve infill pattern were used here for precisely describing the target object. The G-codes of the target object (the nose model) and the printing pattern (a line pattern) were then imported to a custom-written path planning Python programme. In the programme, the z-position of the printing pattern was projected in accordance with the z-position of the target object at the similar x, y positions. By default, the programme assumes that the pattern is printed around the centre of the target object, but an offset position can be used if needed. The programme outputs a text file of the projected coordinate array, which was then imported to the control programme used for Picture input to implement the printing.

**Dispensing of cell suspension**

3T3 mouse embryo fibroblast cell line was cultured in a 25 cm$^2$ flask and was passaged using standard protocol. Cell culture media used here were 10 v/v% fetal bovine serum (F0804, Sigma) and 1 v/v% penicillin-streptomycin (P43333, Sigma) in DMEM (31885023, Life technologies). A cell suspension with 2 x 10$^6$ cells/ml was used in the dispensing experiments, with the cells stained with Calcein AM (C3099, Fisher Scientific) at 2 μM working concentration for live cell staining. To prevent cell sedimentation, immediately after resuspension, the cell ink was drawn into a 1 ml luer-lok syringe and was loaded into the syringe holder of the printer for dispensing operation. The control programme for dispensing operation will be available on Github.

**Ink preparation**

Supplementary Table 4 summarises the inks and the support baths used for fabricating the constructs demonstrated in this work. The inks used here were SE1700 (Dow), 30 w/v% and 40 w/v% Pluronic F127 (P2443, Sigma), a pre-crosslinked alginate ink, a pre-crosslinked hydroxyapatite-alginate ink, 10 w/v% carboxymethyl cellulose sodium salt (21902, Sigma), 10% gelatin (G1890, Sigma), 25% polyacrylic acid (450 kDa, 181285, Merck Life), collagen (50201, Ibidi), a PEGDA solution, 68 wt% methacrylate hydroxypropyl cellulose and 3 w/v% sodium hyaluronate (251770250, Fisher Scientific). Some of the inks were stained with sodium fluorescein (46960, Sigma) or Rhodamine B (A13572.18, Alfa Aesar). Unless further specified, the inks were prepared by dissolving the desired concentration of the chemical powder in deionised water. The methacrylate hydroxypropyl cellulose ink was prepared following the method described in our previous study[39]. The SE 1700 ink was produced by mixing the base precursor and the curing agent at a weight ratio of 10:1. The alginate ink was prepared by pre-crosslinking a 10 w/v% alginate (W201502, Sigma) solution with a 200 mM CaCl$_2$ (C5670, Sigma) solution at a 10:3 volumetric ratio. The hydroxyapatite-alginate ink was made of 15 w/v% hydroxyapatite (21223, Sigma) dispersed in a 5 w/v% alginate solution, which was then pre-crosslinked with a 200 mM $CaCl_2$ solution at a 10:1 volumetric ratio. The PEGDA ink was prepared by mixing PEGDA (Mn 700, 455008, Merck Life), deionised water and a 10 w/v% Irgacure 2959 (g/100 ml ethanol, 410869, Sigma) at a 2:8:1 volumetric ratio. The supportive baths used here were 1.3% xanthan gum, 1 w/v% Carbopol ETD 2020 (Lubrizol) and 4.5 w/v% gelatin slurry. The Carbopol and gelatin slurry supportive baths were prepared following the protocols described in previous studies[40,41].

## Code Availability

All codes for the printing operation will be made available on Github.

## Figures

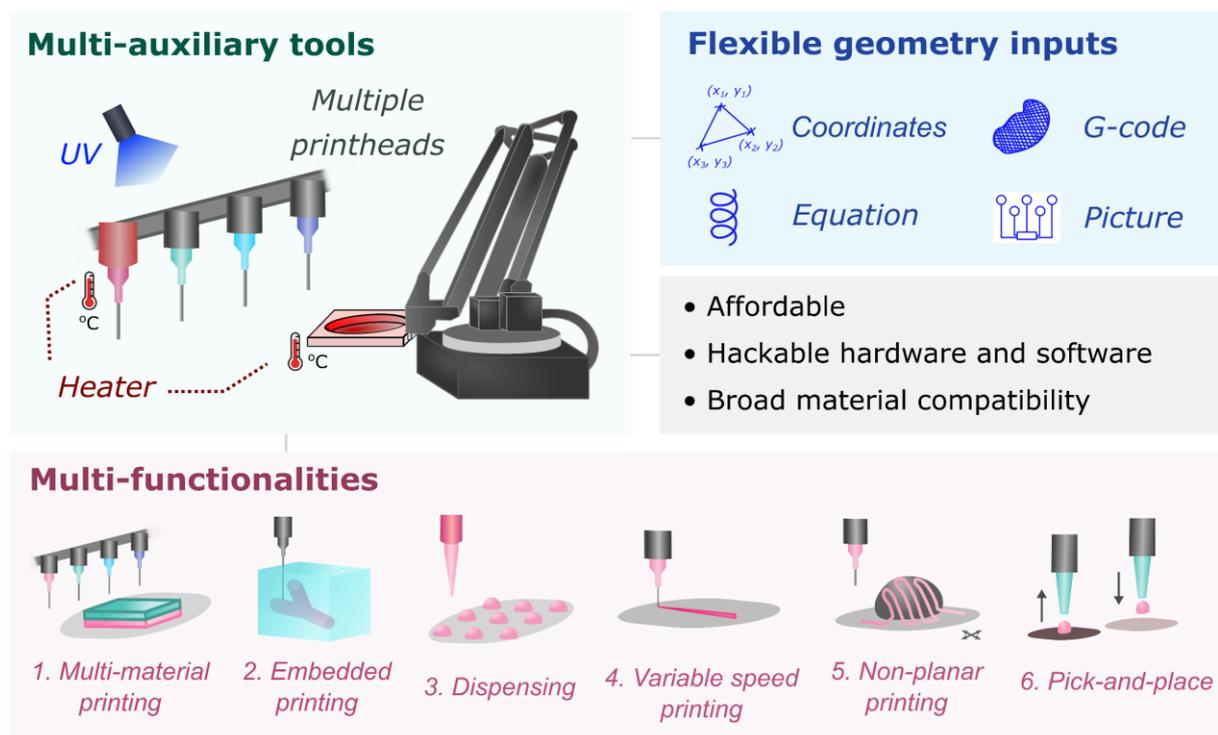

**Figure 1.** Features of Printer.HM, which consists of multi-auxiliary tools as hardware, flexible geometry inputs as software, leading to multi-functionalities in one-platform.

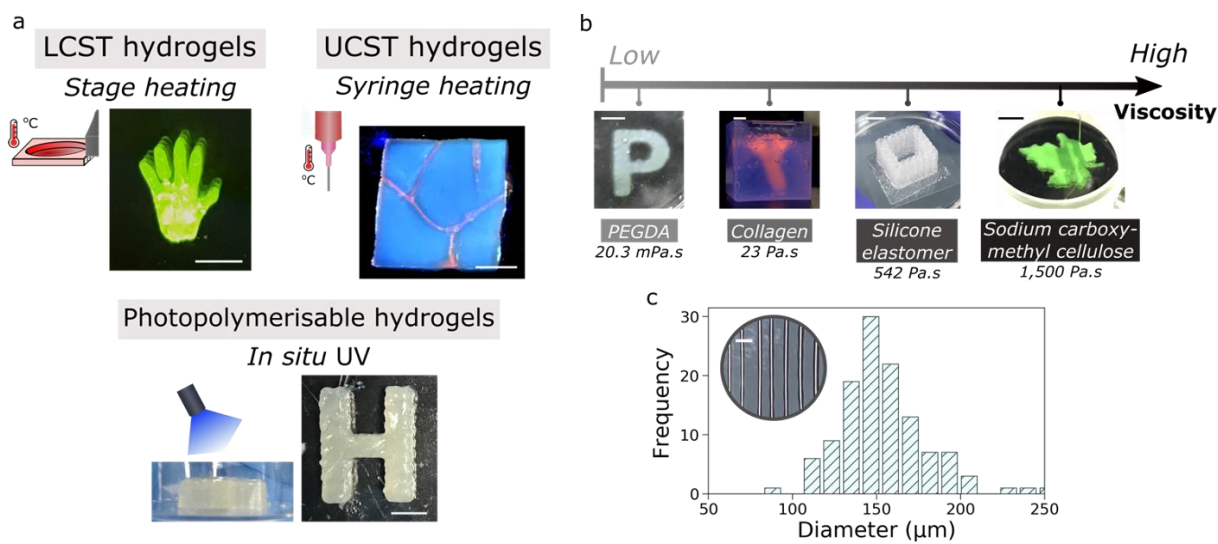

**Figure 2.** a) Figure showing the use of different auxiliary tools for assisting the printing of LCST hydrogels (e.g. Pluronic F127), UCST hydrogels (e.g. gelatin), and photopolymerizable hydrogels (e.g. methacrylate hydroxypropyl cellulose). Scale bar = 5 mm. b) A wide variety of materials over a large viscosity range can be printed using the setup. Scale bar = 5 mm. c) Diameter distribution of the printed Pluronic F127 filaments. Median diameter = ~150 μm. (*n* = 120 measurements over 4 independent samples). Scale bar = 500 μm.

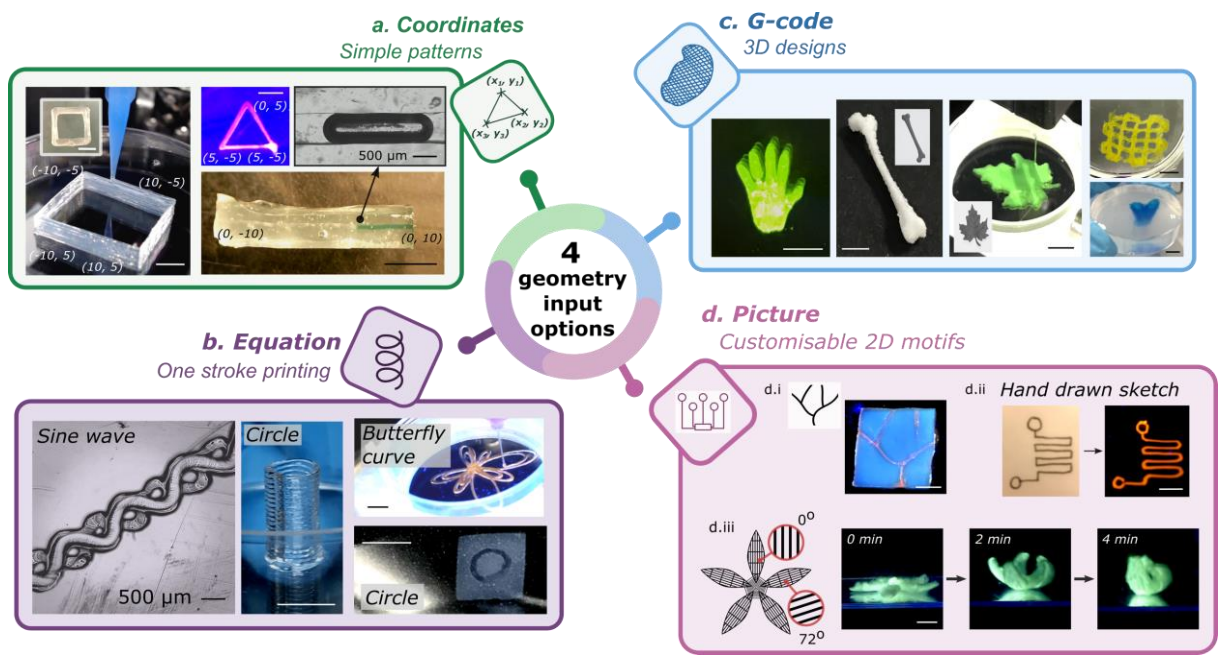

**Figure 3**. Versatile geometry input options enable print creation with different characteristics, via a) coordinates, b) equations, c) CAD models which were then translated into G-code, and d) picture geometries. The materials used here can be found in the Materials and Methods and Supplementary Table 4. Scale bar = 5 mm.

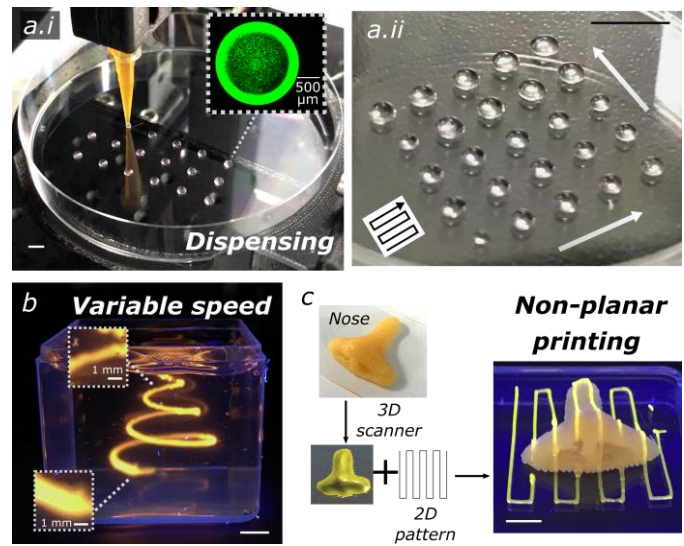

**Figure 4**. Versatile functionalities of Printer.HM. a) Automated dispensing of cell suspension on a petri dish at (a.i) constant droplet volumes and (a.ii) variable droplet volumes. The black arrow in (a.ii) indicates the direction of the dispensing path, and the white arrows depict the controllable droplet size variation from small to large volume. b) A spiral curve made of Pluronic F127 printed with variable speed. c) Non-planar printing of a Pluronic F127 line pattern on a 3D nose model. Scale bars = 5 mm.

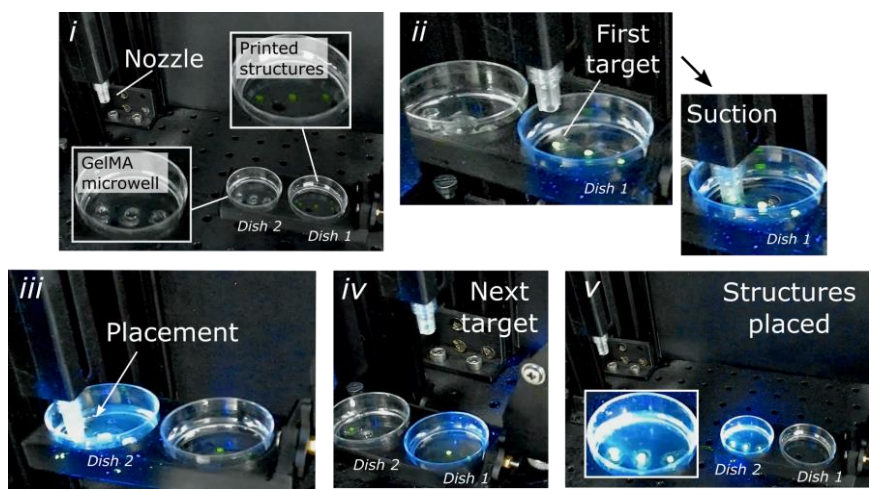

**Figure 5.** Automatic pick-and place-operation, showing i) the pick-up locations of the printed objects and the targeted micro-wells for object placement, ii) Automatic picking process, where 'Dish 1' was translated to the printhead that actuated suction to pick up the structure, iii) Automatic placement process, where 'Dish 2' was translated to the printhead that actuated dispensing to place the structure, iv) Moving to the next target, and v) Completion of the pick-and-place process, where all structures were transferred into the targeted microwells after three iterations of the process. A UV touch was used for illustrating fluorescent structures in ii – v.

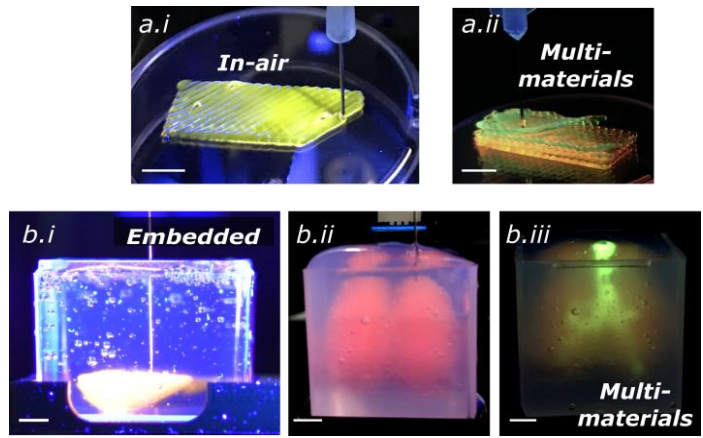

**Figure 6.** Multi-material printing showing a) In-air printing with Pluronic F127 inks stained with different colours, and b) Embedded printing forming a lung structure, with dyed alginate inks printed in a xanthan gum bath.